\documentclass[sigconf,authorversion,nonacm]{acmart}

\usepackage[utf8]{inputenc}
\usepackage[T1]{fontenc}

\usepackage{graphicx}
\usepackage{amsmath}
\usepackage{booktabs}
\usepackage{enumitem}
\usepackage{footmisc}

\renewcommand\footnotetextcopyrightpermission[1]{}

\setcopyright{none}
\acmConference{}
\acmDOI{}
\acmISBN{}

\title{Vibe-Coding: Feedback-Based Automated Verification\\
with no Human Code Inspection, a Feasibility Study}

\author{Michal Töpfer}
\affiliation{%
  \institution{Charles University}
  \city{Prague}
  \country{Czech Republic}
}
\email{michal.topfer@matfyz.cuni.cz}

\author{František Plášil}
\affiliation{%
  \institution{Charles University}
  \city{Prague}
  \country{Czech Republic}
}
\email{frantisek.plasil@matfyz.cuni.cz}

\author{Tomáš Bureš}
\affiliation{%
  \institution{Charles University}
  \city{Prague}
  \country{Czech Republic}
}
\email{tomas.bures@matfyz.cuni.cz}

\author{Petr Hnětynka}
\affiliation{%
  \institution{Charles University}
  \city{Prague}
  \country{Czech Republic}
}
\email{petr.hnetynka@matfyz.cuni.cz}

\acmYear{2026}

\setlist{topsep=0pt}

\begin{document}
\begin{abstract}
Vibe coding inherently assumes iterative refinement of LLM-gene\-rated code
through feedback loops. While effective for conventional software tasks, its reliability in runtime-adaptive systems is
unclear---especially when generated code is not manually inspected.
This paper studies feedback-based automated verification of LLM-generated adaptation managers in Collective Adaptive Systems
(CAS). We focus on the key challenges of verification in the loop: how
to detect failures of generated code at runtime and how to report
them precisely enough for an LLM to fix them.

We combine the adaptation loop with a vibe-coding feedback
loop where correctness is checked against (i) generic architectural
constraints and (ii) functional constraints formalized in Functional
Constraints Logic (FCL), a novel first-order temporal logic over potentially
finite traces. Conducting the Dragon Hunt CAS case study, we show that
fine-grained constraint violations provide actionable feedback that
typically yields a valid adaptation manager within a few iterations,
while simple coarse metric-based feedback often stalls. Our findings suggest that feedback precision is the dominant factor for reliable vibe coding in systems designed by domain experts with no programming skills, thereby obviating the need for human code inspection.
\end{abstract}

\maketitle

\section{INTRODUCTION}

\subsection{Motivation: vibe coding not involving  programmers but domain experts}
Vibe coding has popularized the idea that developers can ``steer''
LLM-generated code by iteratively providing feedback, instead of
reading and editing the code directly \cite{ge2025surveyvibecodinglarge}. Discussions
around vibe coding frequently assume at least light human inspection:
Developers read errors, skim diffs, and reason about fixes. In many practical settings, however, the person who specifies the task is a
domain expert (e.g., game designer, city planner, building engineer)
who cannot reliably inspect code. In that case, the feedback loop
must be fully automated: it must (i) execute the generated artifact,
(ii) detect failures, and (iii) generate failure reports that are
understandable and actionable for the LLM.

Collective Adaptive Systems (CAS) accentuate this challenge. Typically, CAS are complex systems where a large number of heterogeneous agents adapt their behavior to their environment in pursuit of an individual or collective goal. They are controlled by \textit{adaptation manager}, which modifies their settings, assigns them tasks, and organizes them into collaborative groups (also called ensembles) based on a higher-level strategy reflecting the functional goal of each group.
In CAS, correctness is not a purely local property of a function; it
depends on agents' multi-step interactions, dynamic grouping, and runtime
architectural consistency \cite{horizon,horizon2,Muccini}. Thus, a vibe-coded adaptation manager
might compile, run, and still systematically fail because it assigns agents to inappropriate tasks and groups early on, violating the intended
coordination strategy.

\subsection{Setting, key challenge, research question, and contributions}
We study a vibe-coding scenario where:
\begin{itemize}
\item An LLM generates an adaptation manager (AM) that resolves
groups of agents (ensembles) at each adaptation step.
\item The AM is not manually inspected; only feedback-loop refinement
is allowed.
\item The domain expert can state desired behavior as constraints over agents and groups and provide a test set.
\end{itemize}

The key challenge is verification of generated code. 
When the generated AM fails, the system must determine whether the failure is due to (i) a crash or interface mismatch, (ii) an architectural invariant violation (e.g., invalid agent assignment into a group), or (iii) functional misbehavior in a particular group. 
Furthermore, the system must report the failure in the feedback loop precisely enough that an LLM can repair it.

Based on two feasibility studies\footnote{\url{https://github.com/smartarch/llm-adaptation/tree/vibex}} (one of them is described in \autoref{sec:casestudy}), we focus on the following research question:

\smallskip\noindent\textbf{RQ: How efficient is the method of combining the vibe coding feedback loop and the adaptation loop?}

\smallskip
Efficiency is operationalized as the number of feedback iterations required to obtain an AM that passes verification across a set of test executions.

Our approach is based on the hypothesis that feedback granularity is decisive. 
We therefore compare three feedback-level variants: coarse \textit{domain metrics} (baseline), system \textit{generic constraints}, and fine-grained \textit{functional constraints} violations.

\smallskip\noindent\textbf{Overall, the key contributions include:}
\begin{enumerate}
\item A combined adaptation+feedback architecture that turns runtime
verification outcomes into repairable LLM feedback.
\item Functional Constraints Logic (FCL), a novel first-order temporal
logic that supports both infinite and finite traces with explicit trace boundary operators.
\item An empirical evaluation showing that functional constraint-level feedback improves convergence compared to domain metric-level feedback.
\end{enumerate}

\section{CASE STUDY}
\label{sec:casestudy}
\subsection{Dragon Hunt as a CAS}
We use the Dragon Hunt scenario, a fictional game crafted so
that an LLM is unlikely to have been trained to its exact rules. A Dragon
resides in a Cave near a Village. Villagers (agents) are farmers or warriors; they can:
\begin{itemize}
\item Farm wheat in the Village;
\item Spawn new villagers (requires two villagers and sufficient wheat);
\item Move between Village and Cave;
\item Attack the Dragon (more effective for Warriors than Farmers).
\end{itemize}
The Dragon retaliates stochastically, potentially damaging or killing
villagers. The objective is to defeat the Dragon within 30 steps. The scenario is designed so that successful play requires coordinated multi-step strategy: early farming to build an economy, then travel to the cave, and then sustained attacks while managing risk.

\subsection{Ensemble-based modeling and adaptation manager}
We model the Dragoon Hunt scenario as a CAS by using ensemble-based architecture concepts
(similar to DEECo \cite{bures_deeco_2013}). Agents become components---villagers and the dragon; groups of villagers associated with a particular task correspond to ensembles such as Farm, Attack, GoToCave, SpawnFarmer, and SpawnWarrior. At each step, the AM reassigns components to ensembles---performs
\emph{ensemble resolution}:
\begin{enumerate}
\item by observing component attributes (role, location, health) and
environment variables (wheat, dragon HP)
\item assigns each villager to exactly one ensemble, producing the ``architecture'' for that step;
\item applies ensemble effects to advance the system state.
\end{enumerate}
Ensemble resolution is a grouping problem and is NP-hard in general \cite{KOVALYOV20101908}. 
This motivates heuristic synthesis via LLMs and makes the case study a good stress test for vibe coding: local correctness is insufficient; success depends on coordinated sequencing.

\section{VERIFICATION METHOD}
\label{sect.mathod}

\subsection{Combining adaptation and feedback loops}
\begin{figure}[t]
  \centering
  \includegraphics[width=\columnwidth]{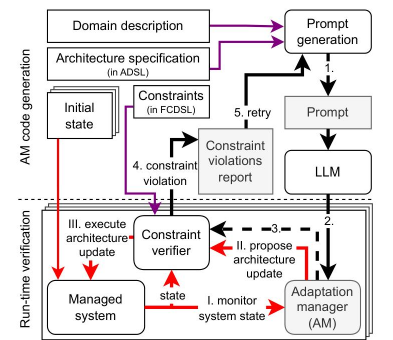}
  \caption{Combining adaptation (red, I.- III.) and feedback (black, 1.-5.) loops.}
  \label{fig:loops}
\end{figure}

Figure~\ref{fig:loops} summarizes the key components of the verification method and their interplay. An AM is generated by an LLM given a prompt based on the domain description and architecture specification. The AM is executed inside the adaptation loop. 
The constraint verifier monitors test runs, and either approves the AM or produces a cumulative constraint-violation report. 
That report is appended to the new version of the prompt, and the LLM is asked to revise the AM.

\subsection{Prompt structure and feedback encoding}
To maximize reproducibility and reduce prompt drift, prompts are
generated from a template. The template includes:
\begin{itemize}
\item \textbf{Interface contract:} required AM functions and return format;
\item \textbf{Domain rules:} scenario description, state variables, available actions/ensembles, effects;
\item \textbf{Strategy intent:} a natural-language description of the desired behavior in terms of ensemble resolution;
\item \textbf{Constraints summary:} optional list of functional constraints in plain language.
\end{itemize}
On failure, the constraint verifier returns a report encoded as a short list of
bullet points, each containing the violated constraint and, optionally, the step window and the most relevant state excerpts. This is intended to mirror how developers communicate bugs: ``what failed'', ``when it failed'', and ``what evidence supports it''.

\subsection{Failure taxonomy in LLM-generated AMs}
In preliminary runs, failures clustered into recurring categories: \textbf{Syntactic/interface failures} include missing functions, wrong return types, invalid imports, or runtime exceptions.
\textbf{Architectural rules failures} include assigning the same villager to multiple ensembles, omitting villagers from any ensemble, or producing unknown ensemble names. 
\textbf{Behavioral/strategy failures} include, despite satisfying architectural rules, still losing because the AM delays 
attacking, mismanages wheat, or violates role/location discipline.
This taxonomy motivates layered constraint verification: generic constraints filter the first two categories, while functional constraints target the third.

\subsection{Verification constraint layers and run coverage}
Verification is performed by executing multiple test runs from varied
initial states and random seeds. A candidate AM is accepted only if it passes all checks across the suite of test runs. The testing is done for the following constraint layers:

\smallskip\noindent\textbf{Generic constraints.}
Importability/executability (no exceptions), valid group/ensemble names, and exactly one group assignment per villager.
These address common LLM failure modes such as missing imports, wrong return shapes, or inconsistent grouping.

\smallskip\noindent\textbf{Functional constraints.}
Requirements expressed in the temporal logic FCL (Section~\ref{sec:fcl}).
These address behavioral misalignment, such as wrong ordering of tasks, starving wheat production, sending farmers to fight, and delaying attacks.

\section{FCL DESCRIPTION AND EXAMPLES OF THE CONSTRAINTS}
\label{sec:fcl}

\subsection{Design goals and semantic intuition}
Functional Constraint Logic (FCL), introduced in \cite{TopeferMPlasilF-FCL}, 
is designed for two goals: (i) expressiveness for bounded multi-step requirements common in CAS strategies, and (ii) producing diagnostic
counterexamples that can be translated into LLM-friendly feedback.
Compared to classical temporal logic LTL, FCL makes trace step counts explicit, which helps in generating informative feedback such as ``attack happens 0 times in steps 1--15''rather than a generic ``globally'', ``sometimes in the future'', ``next'', or ``until'' temporal determination available in LTL.

An execution of a CAS under an AM yields a trace of system states
and ensemble assignments in particular. FCL formulas are evaluated over this trace.
For a given step $i$, the windowed operator $\lozenge^{n}_{t}\varphi$ counts how
many steps in the interval $[i,i+t)$ satisfy $\varphi$ (for $t>0$) and checks
whether the count is at least $n$. This explicit counting makes it
straightforward to produce diagnostic counterexamples: if the count
is below $n$, the verifier can report the exact deficit and list steps
where $\varphi$ was false. For finite traces, \textsc{MAX} and \textsc{BEG}
 (\autoref{subsec:fCLdetails}) allow constraints to ``shrink'' windows near trace boundaries (e.g., ``eventually
before the end''), avoiding spurious failures when a run (actual trace) terminates early.

\subsection{Core operators and finite traces}
\label{subsec:fCLdetails}
The core temporal operator is $\lozenge^{n}_{t}\varphi$, meaning that $\varphi$ holds
at least $n$ times within a window of length $t$ from the current step. A dual ``always'' style
constraint can be encoded by setting $n=t$ 
and past windows are supported by negative $t$. To support finite traces,
FCL includes special counters in the current step:
\begin{itemize}
\item \textsc{BEG}: how many steps from the beginning of the trace;
\item \textsc{MAX}: how many steps remain to the end of the trace.
\end{itemize}
This avoids ambiguous semantics near trace endings and enables
constraints like ``eventually before the end''. There is also the counter \textsc{INF} to support infinite traces.

\subsection{Examples from Dragon Hunt}
\textbf{Win condition.} The Dragon is eventually dead:
\[
\forall d \in \mathit{Dragons} : \lozenge^{1}_{MAX}(\mathit{d.hp} \le 0)
\]
\textbf{Attack early.} The Dragon is attacked at least once within the
first 15 steps:
\[
\lozenge^{1}_{15}(|\mathit{Attack}| \ge 1)
\]
\textbf{Farmers stay in village.}
\[
\forall f \in Farmers : \lozenge^{MAX}_{MAX}(\mathit{f.location} = \mathit{``Village''})
\]
\textbf{Go-to-cave implies eventual attack.}
\[
(v \in \mathit{GoToCave} \wedge MAX>0) \Rightarrow \lozenge^{1}_{MAX}(v \in \mathit{Attack})
\]
\textbf{Economy is not starved.} A simple bounded constraint that
enforces repeated farming early (illustrative):
\[
\lozenge^{3}_{10}(|\mathit{Farm}| \ge 1)
\]
While simplistic, such constraints are useful as scaffolding: they
guide the LLM toward viable strategies, and can later be relaxed or
refined. More examples and details on FCL can be found in \cite{TopeferMPlasilF-FCL}.

\subsection{From counterexample to feedback}
For each violated formula, the constraint verifier constructs a counterexample consisting of the relevant step range, the valuations of referenced sets (e.g., Attack empty), and the smallest witness set (e.g., the specific villager violating a location constraint). 
This is then rendered as a concise textual report. The goal is not to prove properties exhaustively, but to provide a detailed feedback message that supports repair.

\section{EXPERIMENT RESULTS AND VALIDATION}
\subsection{Settings and feedback-level variants}
We implemented a framework\footnote{\url{https://github.com/smartarch/llm-adaptation/tree/vibex}\label{fnlabel}} that realizes all the conceptual components depicted in \autoref{fig:loops} and allows us to control the verification process described in \autoref{sect.mathod}.  Thus, it constructs the prompts and provides input to the LLM through its API,
runs the system adaptation loop with constraint verification, and activates the feedback loop by presenting feedback to the LLM with a constraint violation report. 
For assessing the importance of granularity by which an error is reported in feedback, we have chosen to compare three feedback level variants (listed by refining the granularity of levels): 
\begin{enumerate}
\item \textbf{Metrics-only}: just domain outcome metrics are reported (e.g., win/loss,
remaining Dragon HP, steps survived). This means that no constraints from \autoref{sect.mathod} are considered, so this feedback level serves as a baseline. 
\item \textbf{Generic-only:} only generic constraint failures are reported; functional constraints failures are not explained.

\item \textbf{Full constraint feedback:} both generic and functional constraint
violations are reported. This combination has been chosen since the satisfaction of generic constraints is a natural condition for desired functionality.      

\end{enumerate}
Table~\ref{tab:variants} summarizes the three feedback level variants and the typical
information available to the LLM for code repair.

To illustrate the difference between metrics-only and full constraint feedback consider a run that loses because villagers reach the cave but do not attack.
Metrics-only feedback may only report ``loss, Dragon HP=50'';
the LLM must guess whether to spawn more warriors, change
travel timing, or attack earlier.
Full constraint feedback instead reports
a specific violated obligation, e.g., ``The Dragon should be attacked at least once in the first 15 steps.'' when $\lozenge^{1}_{15}(|Attack| \ge 1)$ is violated.

\subsection{Main result: feedback precision improves convergence}
\begin{figure}[t]
  \centering
  \includegraphics[width=\columnwidth]{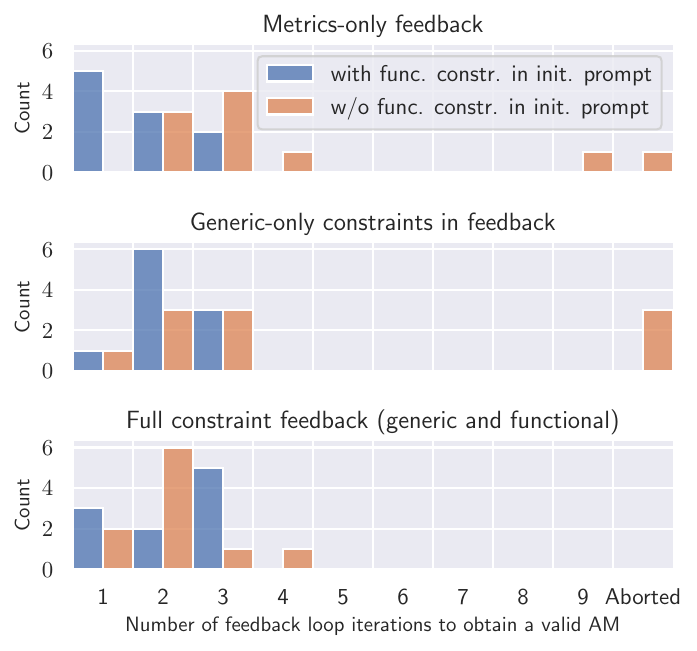}
  \caption{Distributions of feedback-loop iterations needed to obtain a valid AM in the Dragon Hunt example for different feedback variants.}
  \label{fig:dist}
\end{figure}

We measure the efficiency of the method by the number of feedback-loop iterations (\textit{convergence})
needed to obtain an AM that passes verification over the run test suite.
Based on a preliminary study\footref{fnlabel}, a maximum of 10 unsuccessful iterations is
allowed before aborting. 
Figure~\ref{fig:dist} shows the distributions over
10 independent attempts to vibe-code an AM per feedback level variant\footnote{The results in Figure 2 are for GPT 5 nano; furthermore, the results for GPT 5 mini are available at https://github.com/smartarch/llm-adaptation/tree/vibex}.

The results show that full constraint feedback typically converges
within a few iterations. Generic-only feedback is less effective because
many AMs fail functionally but provide no repairable signal. 
Metrics-only feedback often stalls: it indicates failure but does not specify which part of the multi-step strategy is broken. 
This supports the paper's central claim that diagnostic specificity is necessary for vibe coding of CAS featuring multi-step interactions.

With full constraint feedback, same failed attempts are ``one-bug'' iterations: for instance, the AM may satisfy the economy constraints but delay travel to the cave, triggering an early attack violation. The subsequent repair often consists of adding an explicit threshold (e.g., ``if wheat $>$ X then send two warriors to cave'') or prioritizing GoToCave/Attack once the Dragon HP is non-trivial. 

On the other hand, multiple constraints are often violated simultaneously, providing even more precise feedback: ``Warrior John should attack the Dragon after moving to the Cave.'' when the ``Go-to-cave implies eventual attack'' constraint is violated for a particular villager.
In our experience, this tends to elicit focused fixes 
(attack when in a cave)
rather than broad, destabilizing changes. 
In contrast, metric-only feedback frequently leads the LLM to overfit to the metric without understanding the causal structure of the game, producing unstable oscillations such as spawning too many villagers (starving wheat) or sending all villagers to attack
(collapsing the economy).

To further assess robustness, we ensure that the accepted AMs
pass across the whole test suite for multiple initial states and random seeds. Constraint-accepted AMs typically
demonstrate consistent sequencing (farm early, then travel/attack)
across different initial state mixes. This suggests that the constraint
set acts as a strategy-shaped oracle rather than a single-point
performance signal, which is especially important for stochastic domains.

\begin{table}[t]
\caption{Compared feedback variants.}
\label{tab:variants}
\centering
\begin{tabular}{p{0.18\columnwidth}p{0.20\columnwidth}p{0.54\columnwidth}}
\toprule
\textbf{Variant} & \textbf{Oracle} & \textbf{Reporting info (examples)}\\
\midrule
Metrics & outcome score & win/loss; remaining Dragon HP; steps survived; wheat at end\\
Generic-only & execution + arch. rules & exception trace; ``invalid ensemble name''; ``component assigned twice''\\
Full constraint & arch. rules + FCL & violated formula; time window; witness agents/sets, e.g., ``No attack in steps 1--15''\\
\bottomrule
\end{tabular}
\end{table}
\subsection{Validation and interpretation}
A key concern is whether ``passing constraints'' corresponds to genuinely
correct behavior. In our method, constraints are treated as
the formalization of the intended strategy, and the test suite is
designed to achieve high coverage over representative system states. In this sense, constraints
act as executable requirements: they operationalize the domain expert's
intent. Nevertheless, we interpret the results as feasibility evidence rather
than a definitive guarantee: the approach is only as good as the constraint set
and run coverage.

\section{DISCUSSION AND RELATED WORK}

\subsection{Relation to vibe coding feedback loops and formal verification for LLM-generated code}
Most vibe coding tooling focuses on compilation errors, unit tests, static analyzers, and even fuzzy testing \cite{nunez2024autosafecodermultiagentframeworksecuring,ge2025surveyvibecodinglarge}. 
These are highly effective for programs with local properties and clear oracles. 
Our setting differs: the oracle is temporal, multi-step, and architectural. We show that integrating runtime temporal constraints is a practical way to extend vibe coding beyond unit-test-centric tasks.

There is growing interest in bringing formal methods to LLM outputs, including verification and runtime monitoring \cite{councilman2025formalverificationllmgeneratedcode,zhang2025rvllmllmruntimeverification}. 
Our contribution is complementary: we target bounded temporal requirements and explicitly aim to produce repairable feedback, not only verdicts. 
From a workflow perspective, the verifier plays the role of an automated ``reviewer'' that points to specific violations.

\subsection{Self-adaptation and collective systems context}

Our setting is motivated by a long line of work on self-adaptive systems \cite{Tpfer2025,bures_deeco_2013}. 
In these systems, adaptation logic is traditionally engineered and validated using model-based or
rule-based techniques. 
Recent surveys highlight growing interest in employing generative AI in self-adaptive systems \cite{Zhong2024}. 
Compared to approaches that embed ML components inside the runtime loop, our work focuses on using LLMs to synthesize the control logic itself and then validating it through repeated executions plus constraint monitoring.
This aligns with calls for rigorous engineering of collective adaptive systems \cite{Denicola}.

\subsection{Why temporal constraints fit vibe coding and implications for VibeX practitioners}

A recurring pattern in vibe coding studies is that improvements come from turning vague failures into precise, localized signals (e.g., failing unit tests, minimized repro cases \cite{mundler2025swtbench}). 
In multi-agent and collective adaptive systems, the natural oracles are temporal and relational: the failure may be ``nothing happens for too long'' or ``the wrong agents coordinate''.
Windowed temporal constraints like those in FCL provide exactly this kind of localized signal: they identify when an obligation was missed and which agents or ensembles are involved. 
This allows the LLM to remain in a narrow repair mode rather than triggering broad rewrites.

For practitioners using vibe coding in domains with runtime interaction (games, simulations, robotics, distributed coordination), our results suggest two practical guidelines. 
First, invest in constraint instrumentation early: even a small set of temporal constraints can dramatically improve feedback quality compared to scalar metrics. 
Second, treat constraint sets as evolving artifacts: begin with strategy-shaped scaffolding constraints (e.g., ``attack early'') and refine them as the loop stabilizes. 
This mirrors test-driven development but extends it to temporal requirements, bringing a ``requirements-as-tests'' mindset to vibe coding.

\subsection{Future work and threats to validity}

Two directions appear particularly promising. 
First, constraints can be used not only for monitoring but also for guidance during generation: the prompt could include machine-readable constraint fragments or automatically derived intermediate goals. 
Second, the verification suite could be expanded from deterministic runs to statistical coverage, combining temporal constraints with randomized scenario generation to systematically surface rare but harmful failures. 
Both directions aim to reduce the manual effort of ``constraint engineering'' while increasing confidence in automated vibe coding.

Construct validity: constraints may mis-specify intent. 
Internal validity: LLM and environment stochasticity; mitigated by repetition and controlled seeds. 
External validity: Dragon Hunt is synthetic; larger CAS may require richer constraints and scalable execution. 
Reliability: prompt sensitivity; template-based prompting reduces variability but does not eliminate it.

\section{CONCLUSION}
We studied feedback-based automated verification for vibe-coded adaptation managers in CAS under a no-inspection assumption. 
By integrating runtime verification into the feedback loop and expressing functional requirements as fine-grained temporal constraints in FCL, the feedback loop converges efficiently in most runs.
Coarse metric feedback frequently stalls because it lacks a causal diagnostic signal.

For VibeX, the key takeaway is that, for runtime-dependent adaptive systems, the feedback structure matters more than the iteration count: temporal constraint violations transform refinement from trial-and-error into directed correction.

\vspace{1mm}
\noindent\textbf{Acknowledgment:} This work was partially supported by the EU HORIZON-JU-Chips project NexTArc (EU grant agreement No. 101194287, Czech national funding Id 9A25008) and partially by Charles University institutional funding 260821.

\bibliographystyle{ACM-Reference-Format}
\bibliography{paper}

\end{document}